\documentclass[prb,twocolumn,showpacs,preprintnumbers,amsmath,amssymb]{revtex4}
\usepackage{graphicx, epsfig}
\usepackage{dcolumn}
\def\be{\begin{equation}}
\def\ee{\end{equation}}
\def\bea{\begin{eqnarray}}
\def\eea{\end{eqnarray}}

\begin{document}
\title{Topological and topological-electronic correlations in amorphous silicon}
\author{Yue Pan}\email{pan@phy.ohiou.edu}
\author{Mingliang Zhang}\email{zhang@phy.ohiou.edu}
\author{D. A. Drabold}\email{drabold@ohio.edu}
\affiliation{Department of Physics and Astronomy, Ohio University, Athens, OH 45701}
\date{\today }

\begin{abstract}
In this paper, we study several structural models of amorphous silicon, and
discuss structural and electronic features common to all. We note spatial
correlations between short bonds, and similar correlations between long
bonds. Such effects persist under a first principles relaxation of the
system and at finite temperature. Next we explore the nature of the band
tail states and find the states to possess a filamentary structure. We
detail correlations between local geometry and the band tails.
\end{abstract}
\maketitle

\section{Introduction}

The nature of three-dimensional space-filling disordered networks is salient to
a variety of topics in scientific research. Despite work in recent decades, a number of puzzles
remain both about the structure and connectivity of such networks. Also, important
quantities {\it derived from} such structures (such as electronic or vibrational states) are
quite incompletely understood, though their characteristics are ultimately derivative from
the underlying structure. In this paper, we systematically consider a collection of
(largely) tetrahedral continuous random networks of amorphous Si, and
and despite the fact that several of these are quite ``standard", we
detect and report novel features of their connectivity and electronic structure.

The character of the tail states in disordered semiconductors is a problem
of importance with a history dating back at least to the sixties. It has
long been understood that band tail states in amorphous
semiconductors arise from strains in the network sufficient to push states
past the band edges into the gap. Such tails decay exponentially into the
gap: this feature, nearly universal to disordered semiconductors,
is called Urbach tailing. \cite{urbach} Clever
photoemission experiments allow the separate observation of the valence and
conduction tails and even the temperature dependence of the tailing \cite{aljishi}. 
Several theories have been presented to explain the Urbach tails, most recently by Cohen and coworkers. \cite{morrellcohen,cohen2}

Beside band tailing and the detailed structure of the
networks we study, our work is related to the general problem of
the topology of space-filling networks. For example, work on the cuprates \cite{jcp,jcp89} provides
clear, albeit circumstantial evidence for the existence of percolating filamentary networks, and is seen in STM measurements \cite {davis} (presumably two-D projections of the three-D filamentary network). Filamentary paths are also believed to exist in fast ion conducting glasses with larger than average free volume \cite{swenson} regions providing pathways for ion transport (see for example Fig. 3 in Ref. \cite {swenson}). Farther afield, results akin to ours (the percolation of short bonds) can be inferred in protein binding in a yeast \cite{pnas}.

This paper has two immediate aims. First, consideration of several models reveals
certain persistent geometrical features, notably strong self-correlations
between configurations involving short bonds and analogous correlations
involving long bonds. To the extent that these correlations exist in various
structural models and survive \textit{ab initio} thermal molecular dynamics
(MD) simulations, it seems worthwhile to point these out. The second aim is
to discuss general features of the band tail states in a-Si. Resonant
cluster proliferation of tail and gap states is observed \cite{Resonance},
and filamentary connections between \ localization centers. In agreement
with earlier reports \cite{Fedders,Kang}, we observe a robust tendency for
valence (conduction) tail states to be associated with short (long) bonds.
Unlike our earlier report \cite{Fedders}, we did detect analogous features
for the bond angle distribution, where we see that smaller (larger) bond
angles are correlated with valence (conduction) tails.

These general observations about the tail states in a-Si are relevant to
calculations of transport, since it is presumably just these states through
which carriers hop, as we discuss in detail elsewhere \cite{kuboprb}.

\section{Constraints on Theory}

\subsection{Model characteristics and nomenclature}

Even in good quality unhydrogenated material, mid-gap defects are fairly
rare, at most a few sites per thousand \cite{street}. At a practical level
this means that ideal models of tractable size (up to several thousand
atoms) can contain zero to a few defects producing mid-gap states. The
Wooten-Weaire-Winer (WWW) \cite{www} method is the \textquotedblleft gold
standard\textquotedblright\ for forming such realistic models, and several
of the models we discuss depend upon WWW at least as a starting point. For
completeness, and in particular to ascertain the extent to which our
observations might tend to be WWW artifacts, we have repeated the
calculations for other models as well. We summarize some salient features of
the models in Table I.

\begin{tabular}{|l|l|l|l|l|l|l|}
\hline
\# & Author & NOA & Type & $\Delta \theta $/$\Delta \cos \theta $ & $\Delta
b $ & De\% \\ \hline
M$_{1}$ & DTW \cite {DTW} & 512 & WWW & 20.1$^{\circ }$/0.33 & 0.08 & 0.0 \\
M$_{2}$ & DTW \cite {DTW} & 4096 & WWW & 19.1$^{\circ }$/0.31 & 0.19 & 0.2 \\
M$_{3}$ & Feldman \cite {Feldman} & 1000 & WWW & 18.4$^{\circ }$/0.30 & 0.08
& 4.0 \\
M$_{4}$ & Mousseau \cite {art} & 4000 & ART\cite {art} & 17.8$^{\circ }$/0.30
& 0.11 & 4.0 \\
M$_{5}$ & Nakhmanson \cite {serge} & 1000 & PC & N/A & N/A & 0.0 \\
M$_{6}$ & Biswas \cite {rmc} & 216 & RMC\cite {rmc} & 22.3$^{\circ }$/0.37 &
0.30 & 12.5 \\
M$_{7}$ & Biswas \cite {rmc} & 500 & RMC\cite {rmc} & 22.5$^{\circ }$/0.37 &
0.24 & 12 \\
M$_{8}$ & Mousseau & 64 & WWW & 20.7$^{\circ }$/0.34 & 0.11 & 0.0 \\
M$_{9}$ & DTW \cite {DTW} & 216 & WWW & 24.5$^{\circ }$/0.40 & 0.30 & 0.0 \\
\hline
\end{tabular}

TABLE 1. Models and basic information. NOA stands for \textquotedblleft
Number of atoms\textquotedblright ; De\% stands for \textquotedblleft
defects percentage\textquotedblright ; $\Delta \theta $/$\Delta \cos \theta $
stands for the width of bond-angle $\theta $/$\cos \theta $ distribution; $%
\Delta b$\ stands for the width of bond-length distribution; WWW stands for
Wooten-Weaire-Winer modeling scheme\cite{www}; ART stands for
activation-relaxation technique\cite{art}; PC stands for \textquotedblleft
paracrystalline\textquotedblright ; RMC stands for Reverse Monte-Carlo
modeling scheme\cite{rmc}.

M$_{1}$, M$_{2}$, M$_{3}$, M$_{8}$ and M$_{9}$ all derived originally from
WWW \cite{www} modeling scheme. M$_{1}$, M$_{2}$ and M$_{9}$
were developed by B. R. Djordjevic, M. F. Thorpe, F. Wooten \cite{DTW}; M$%
_{3} $ was developed by Feldman and M$_{8}$ by N. Mousseau. M$_{4}$ was made
via activation-relaxation technique \cite {art}. M$_{5}$ and M$_{6}$ are
Reverse-Monte-Carlo \cite{rmc} models. Model M$_{1}$ to M$_{4}$ and M$_{6}$
to M$_{9}$ are continuous random network (CRN) models. Beside CRN models,
we include model M$_{5}$, a paracrystalline\ model (CRN with crystalline
inclusion). In particular for M$_{5}$, 211 crystalline silicon atoms are
embedded.

\begin{figure}
\begin{center}
\resizebox{90mm}{!}{\includegraphics{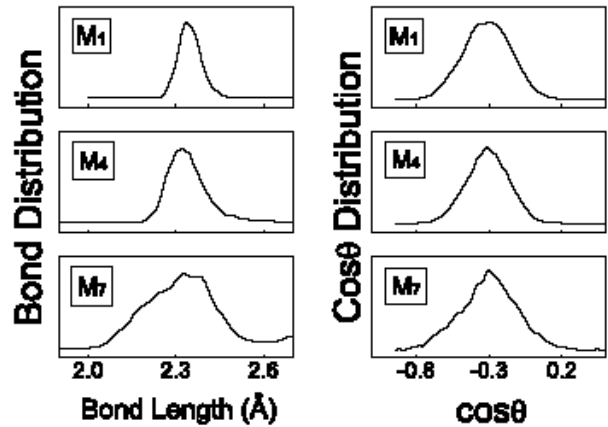}}\end{center}
\caption{Bond length distribution for model M$_{1}$, M$_{4}$ and M$_{7}$
(left column); Cosine bond angle distribution for model M$_{1}$, M$_{4}$ and M$_{7}$
(right column).}
\label{F01}
\end{figure}

Excepting M$_{5}$, all the models that we have studied here have normally
distributed bond lengths and cosines of bond angles. For the large WWW
models the similarity to the normal distribution is quite striking \cite{dong96}. Here we report the bond length and bond angle distributions for
these models in Fig. \ref{F01}. Model M$_{1}$ and M$_{4}$ show nearly
perfect Gaussian fits, while model M$_{7}$, (a 500-atom model), is fairly
Gaussian, with statistical noise. The other models exhibit similar traits. Dong and Drabold \cite{dong96} have attempted to correlate these normally-distributed quantities to Urbach tailing.

Model M$_{1}$, M$_{5}$, M$_{8}$ and M$_{9}$ are defect free. In the other
models, there are coordination defects, ranging from 0.2\% to 12.5\%,
including both dangling bonds and five-fold \textquotedblleft
floating\textquotedblright\ bonds. The pair correlation functions of these
models are all similar, accurately reflecting the experimental function. As
usual with a-Si, this is a reminder that the information in the
pair-correlation function alone is very insufficient to specify coordinates.

\subsection{Local inter-bond correlations}

\begin{figure}
\begin{center}
{\includegraphics[angle=0, width=0.5\textwidth,height=0.35\textheight]{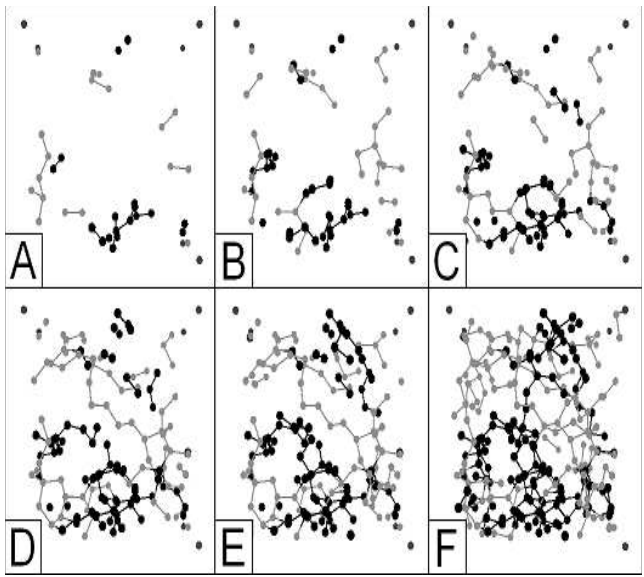}}
\end{center}
\caption{(A) 1\%, (B) 2\%, (C) 3\%, (D) 4\%, (E) 5\% and (F)8\% shortest(dark) and
longest(light) bonds of model M$_{1}$.}
\label{F02}
\end{figure}

\begin{figure}
\begin{center}
{\includegraphics[angle=0, width=0.52\textwidth,height=0.35\textheight]{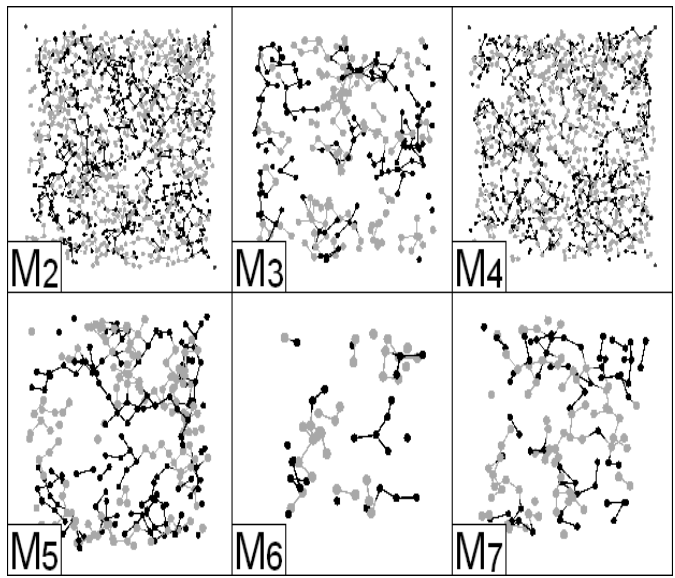}}\end{center}
\caption{4\% shortest(dark) and longest(light) bonds of model M$_{2}$, M$%
_{3} $, M$_{4}$, M$_{5}$, M$_{6}$ and M$_{7}$.}
\label{F03}
\end{figure}

Our inspection of the topology of the models reveals that there is a
tendency for short bonds to be linked to other short bonds and long bonds to
be linked to other long bonds. In particular, this leads to 1) a tendency
to spatial separation of the shortest and the longest bonds in the
system; 2) a tendency to build up chain and ring like structures among
extreme bonds. Selected fractions of shortest and longest bonds of model M$%
_{1}$ are extracted from the models and displayed in Fig. \ref{F02}.
The correlation is clearly visible, especially in Fig. \ref{F02}c-f.
Interpenetrating filamentary structures appear to percolate through space
for a sufficiently large cutoff in the fractions of bonds displayed. For
comparison, the 4\% shortest and 4\% longest bonds of model M$_{2}$ - M$_{7}$
are shown in Fig. \ref{F03}. To some extent, the self-correlations of short
and long bonds are visible in every model shown, suggesting that such bond
correlations are not an artifact of a particular model or modeling scheme.

\begin{figure}
\begin{center}
\resizebox{90mm}{!}
{\includegraphics{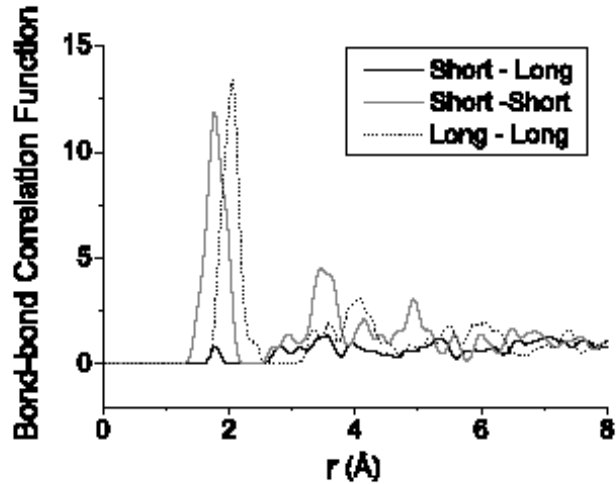}}\end{center}
\caption{Bond-bond correlation function between selected bonds of model M$%
_{1}$. Solid black line denotes pair correlation function between the 4\%
shortest and the 4\% longest bonds. Grey line denotes pair correlation
function among the 4\% shortest bonds. Dotted line denotes pair correlation
function among the 4\% longest bonds.}
\label{F04}
\end{figure}

\begin{figure}
\begin{center}
\resizebox{90mm}{!}
{\includegraphics{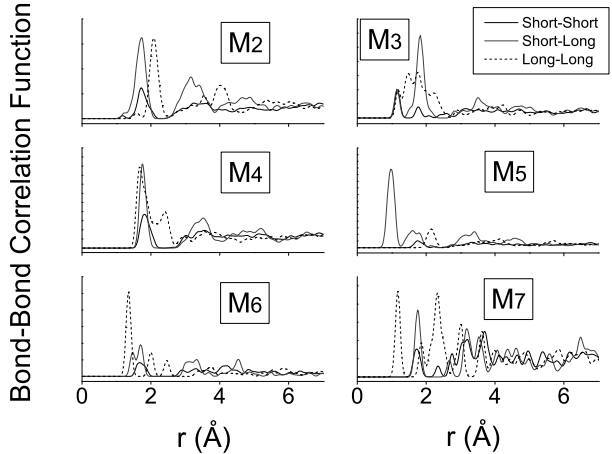}}\end{center}
\caption{Bond-bond correlation functions between selected bonds of model M$%
_{2}$, M$_{3}$, M$_{4}$, M$_{5}$, M$_{6}$ and M$_{7}$. Solid black line
denotes pair correlation function between the 4\% shortest and the 4\%
longest bonds. Grey line denotes pair correlation function among the 4\%
shortest bonds. Dotted line denotes pair correlation function among the 4\%
longest bonds.}
\label{F05}
\end{figure}

To explore our observation more quantitatively, we computed bond-bond
correlation functions for the shortest and longest bonds:
\begin{equation}
\bigskip \beta (r)=\frac{V}{4\pi r^{2}N_{1}N_{2}}\sum%
\limits_{n_{1}=1}^{N_{1}}\sum\limits_{n_{2}\neq n_{1}}^{N_{2}}\delta
(r_{n_{1}n_{2}}-r)  \label{EQ00}
\end{equation}%
where $N_{1}$, $N_{2}$ are the number of short or long bonds as needed; $%
n_{1}$ and $n_{2}$ count over these particular bonds; $r_{n_{1}n_{2}}$\ is
the distance between the bond centers of bond $n_{1}$ and bond $n_{2}$; $V$
is the volume of the unit cell. The 4\% shortest and 4\% longest bonds are
taken for model M$_{1}$ and the corresponding bond-bond correlation
functions for (1) $N_{1}$ = 4\% shortest, $N_{2}$ = 4\% longest; (2)$N_{1}$
= 4\% shortest, $N_{2}$ = 4\% shortest; (3) $N_{1}$ = 4\% longest, $N_{2}$ =
4\% longest are plotted in Fig. \ref{F04}. The first peaks of the
correlations of the same-type bonds (grey and dotted lines) appear to be
much stronger than the correlation between the different-type bonds (solid
black line). The second and even the third peaks maintain this feature, and
after 6.0 $\mathring{A}$, this correlation wanes. This confirms that there is a
correlation among the same-type-bonds, with a correlation radius around 6.0$%
\mathring{A}$ for model M$_{1}$. The very similar correlation functions
between the 4\% shortest and the 4\% longest bonds of model M$_{2}$ - M$_{7}$
are plotted in Fig. \ref{F05}. A substantial correlation between same-type
bonds is revealed for all models, with minor variations from model to model.
Model M$_{1}$, M$_{3}$, M$_{5}$ and M$_{6}$ show slightly stronger effects
than the rest. Model M$_{3}$ and M$_{5}$ show stronger correlations between
the short bonds compared to those between the long bonds, while model M$_{6}$
and M$_{7}$ have relatively stronger correlations between the long bonds
than those between the short bonds.

\subsection{Persistence of correlation effects with \textit{ab initio}
relaxation}

The correlations are found quite consistently in the several models we
examined, suggesting that the effects are not modeling artifacts. We also
undertook a simple comparison to see how a conjugate gradient (CG)
relaxation process by SIESTA affects the models. A relaxation was done on
model M$_{1}$ and M$_{9}$. While there were slight modifications in
structure due to the CG process, we didn't observed any significant changes
in these correlations.

\subsection{Persistence of correlation at finite temperature}

\begin{figure}
\begin{center}
\resizebox{90mm}{!}
{\includegraphics{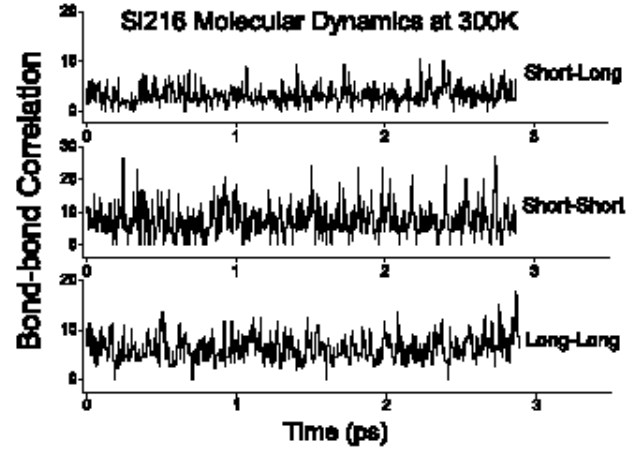}}\end{center}
\caption{Correlation (1) between the 4\% shortest bonds and the 4\% longest
bonds; (2) among the 4\% shortest bonds; (3) among the 4\% longest bonds of
model M$_{9}$ throughout time steps at 300K. Nos\'{e} dynamics were used.}
 \label{F06}
\end{figure}

To further explore the likelihood that the long and short inter bond
correlations were artifacts, we also undertook thermal MD simulations at
300K on two small models (M$_{8}$ and M$_{9}$) with the code SIESTA. Nos\'{e}
dynamics were used. With some thermally-induced fluctuations, the
short-short bond correlation and long-long bond correlation persist at all
times, as seen in Fig. \ref{F06} for model M$_{9}$, where the heights of the
first peaks of the pair correlation functions from Eq. (\ref{EQ00}) are
plotted throughout time steps. 4\% shortest bonds and 4\% longest bonds of
the system at all time were involved for the calculations. The correlations
between the same types of bonds are fluctuating but generally higher than
that between the different types. The detailed temperature dependence in
this correlation, however, requires further study.

\section{Electronic Structure}

\subsection{Density of states and Localization}

The electronic and optical properties of the models are determined by the
coordinates of the atoms. \ Electronic eigenstates and wavefunctions of the
models are obtained for static lattices. For the larger model M$_{2}$ and M$%
_{4}$ a tight binding model\cite{TB,tbmd} is applied. For other models, the
\textit{ab initio} code SIESTA \cite{Siesta} is used. Results are very
similar for the empirical and \textit{ab initio} calculations.

Densities of states near E$_{f}$ of the seven models are given in Fig.\ref%
{F07}. Spatial localization is reported using the Inverse Participation
Ratio (IPR) in Fig. \ref{F08}. Model M$_{1}$ and M$_{5}$ are defect free,
with no states in their optical gaps. Model M$_{2}$ has 0.2\% defects and
one gap state. Model M$_{3}$ and M$_{4}$ both have 4\% defects, and a small
peak of states in the gap. M$_{6}$ has 12.5\% defects and M$_{7}$ has 12\%
defects and both have many states spread through the gap. Only M$_{1}$, M$%
_{2}$ and M$_{5}$ conform closely to ordinary expectations for an a-Si DOS. M%
$_{3}$, M$_{4}$, M$_{6}$ and M$_{7}$ exhibit too many defects for good a-Si
materials and are studied here mainly for comparison purpose.

\begin{figure}
\begin{center}
\resizebox{90mm}{!}
{\includegraphics{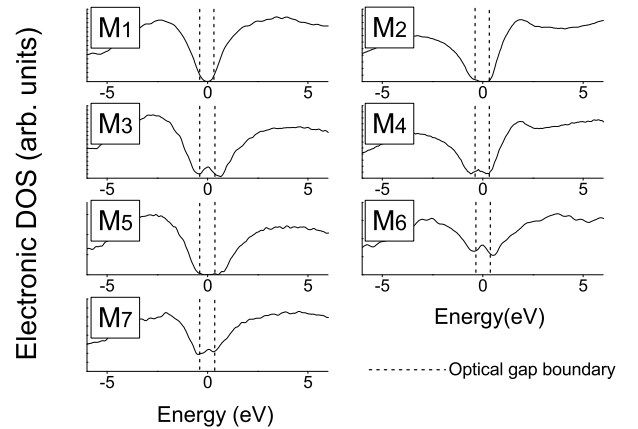}}\end{center}
\caption{Densities of states of models. Dashed lines indicate optical gaps.
Fermi levels are all shifted to 0.0 eV.}
\label{F07}
\end{figure}

\begin{figure}
\begin{center}
\resizebox{90mm}{!}{\includegraphics{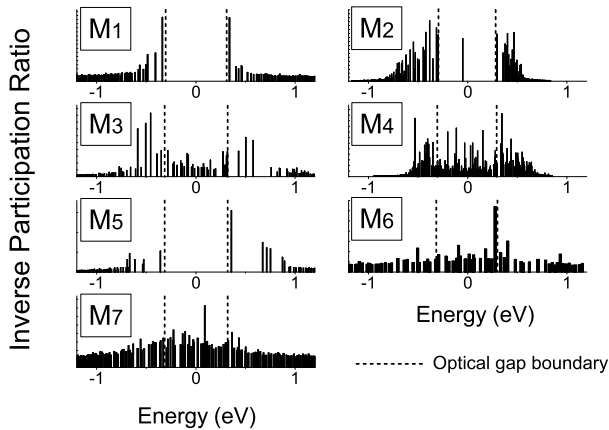}}
\end{center}
\caption{Spatial localization obtained from IPRs (Inverse Participation
Ratio) of models. Dashed lines indicate optical gaps. Fermi levels are all
shifted to 0.0 eV. For M$_{2}$ and M$_{4}$ only a small part of IPRs near
the Fermi level is shown.}
\label{F08}
\end{figure}

\subsection{Topology of band tail states}

\subsubsection{Filamentary structure}

\begin{figure}
\begin{center}
\resizebox{90mm}{!}
{\includegraphics{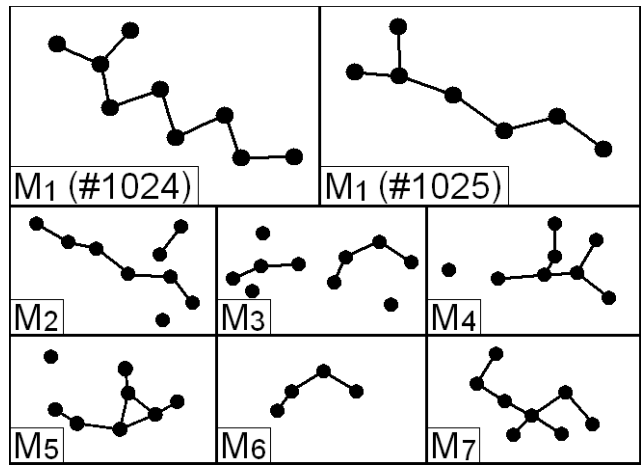}}\end{center}
\caption{Most localized electronic eigenstates of model M$_{1}$, M$_{2}$, M$%
_{3}$, M$_{4}$, M$_{5}$, M$_{6}$ and M$_{7}$. Approximately 60\% - 75\%
charge of the state is shown in each case. For model M$_{1}$, both a valence
tail state(\#1024) and a conduction tail state(\#1025) are shown.}
\label{F09}
\end{figure}

The most-localized states of model M$_{1}$, displays a 1-D filament connectivity.
For state \#1024(-0.34eV, a valence state) and State \#1025(0.34eV, a
conduction state) the most electron-probable atoms are selected and shown
in Fig. \ref{F09}. Interestingly, in both case, the atoms locate closely
(localization) and happen to form filament (chain) like structures. In
addition, the angles along the chain of E\#1024 are all small angles, and
the bonds are all short bonds. The angles along the chain of E\#1025 are all
large, and the bonds all long.

Localized states of the other six models, M$_{2}$ to M$_{7}$, were also
examined. We picked the most localized state for each model. Only the atoms
possessing most charge are shown in Fig. \ref{F09}. We see that, for all
these models, these localized atoms form filament-like structures in
general. The strength of the effect varies only slightly among the models.

\subsubsection{Bond length-electron correlations and bond angle-electron
correlations near E$_{f}$}

Fedders, Drabold, Nakhmanson's study \cite{Fedders} observed that valence
tail states preferentially involved short bond lengths, whereas conduction
tail states tended to be involved with longer bond lengths. Angle distortion
was also mentioned but no specific relation to the band tails was detected.
Related work was undertaken in Ref. \cite{Kang}. Here, we extend this
analysis to our collection of models. Where the bond lengths are concerned,
the present work is consistent with these findings. In addition we also
clearly detect a correlation between valence tail and small bond angle, and
a correlation between conduction tail and large bond angle, which has not
been explicitly observed before.

An electron(Mulliken charge)-weighted mean bond length for each state was
calculated in Ref. \cite{Fedders}, and it revealed an interesting and
general trend in bond lengths. Here we computed the mean bond length in a
similar fashion, and explored into seven of our models for this mean value.
We then extended the calculations into the mean bond angles, using the very
same Mulliken charge.

\begin{figure}
\begin{center}
\resizebox{90mm}{!}
{\includegraphics{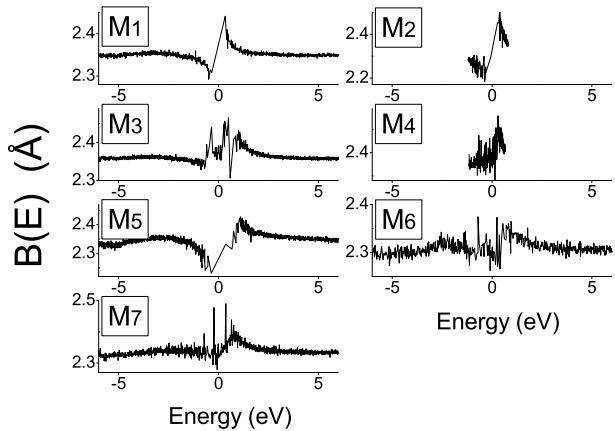}}\end{center}
\caption{B(E), the electron-weighted average of bond length near the Fermi
levels for the seven models. Only small fractions of data are showing for M$_{2}$
and M$_{4}$.}
 \label{F10}
\end{figure}

For a certain eigenstate $E$, the Mulliken charge $q_{(n,E)}$ is the
electron probability on an atom $n$. Using it as a weighting factor, we
define the corresponding symmetrized mean bond length as $B_{(E)}$.
\begin{equation}
B_{(E)}=\frac{\sum_{n,m}b_{(n,m)}q_{(n,E)}q_{(m,E)}}{%
\sum_{n,m}q_{(n,E)}q_{(m,E)}}  \label{EQ02}
\end{equation}%
where $n,m$ summations \textit{only} go over all the possible bonds in a
unit cell (of $N$ atoms), with $n,m$ as the bond end atoms. $q_{(n,E)}$, as
defined, is the electron probability on the atom n. $B_{(E)}$ (near the
Fermi level) for each model is plotted in Fig. \ref{F10}. $B_{(E)}$ shows an
asymmetrical split around the gap (Fermi level) for all models. Smaller
average bond lengths are correlated to the valence tails and large average
bond lengths correlated to the conduction tails. There are clear difference
among the seven models, especially on the mid-gap states from the models with
high defect concentration. For models with fewer defect states in the gap,
the bond-electron correlations are very clear. Models with many mid-gap
states show mixed behavior, though the asymmetric splitting is always
detectable.

\begin{figure}
\begin{center}
\resizebox{90mm}{!}
{\includegraphics{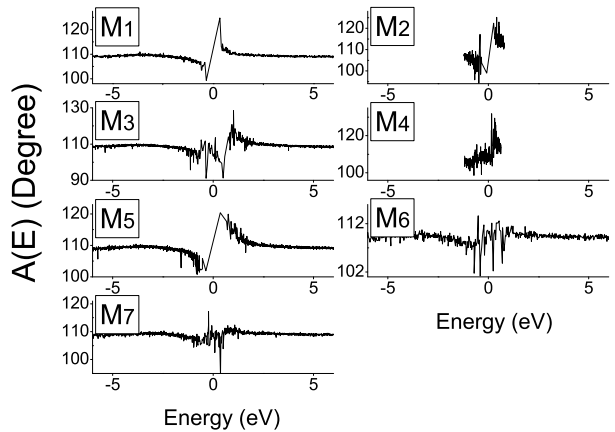}}\end{center}
\caption{A(E), the electron-weighted average of bond angle near the Fermi
levels for the seven models. Only small fractions of data are showing for M$_{2}$
and M$_{4}$.}
 \label{F11}
\end{figure}

In analogy with the discussion for bond lengths, we define the
electron-weighted mean of bond angles as:
\begin{equation}
A_{(E)}=\frac{\sum_{n,m,l}\theta _{(n,m,l)}q_{(n,E)}q_{(m,E)}q_{(l,E)}}{%
\sum_{n,m,l}q_{(n,E)}q_{(m,E)}q_{(l,E)}}  \label{EQ03}
\end{equation}%
where $n,m,l$ are the atoms of the unit cell of $N$ atoms; summations of n,
m, l \textit{only} count over all the possible bond angles, with m as the
vertex of each angle. $q_{(n,E)}$, as defined, is the electron probability
on the atom n. $A_{(E)}$ is plotted in Fig. \ref{F11}. It also
shows a clear asymmetrical splitting around the gap for all plots. Smaller
average bond angles clearly appear to be correlated to the valence tails and
larger average bond angles correlated to the conduction tails. There are
also differences among the models but the asymmetric feature remains.
Interestingly, these angle-electron correlations show strong resemblance to
the bond-electron correlations discussed above. It is also natural to
connect these correlations to the filamentary structures shown in Fig. \ref%
{F09}. For M$_{1}$, which is a typical example, the structure from state
\#1024 (a valence tail state) has both short bonds and small bond angles
along the chain, while \#1025 (a conduction tail state) has both long bonds
and large bond angles along the chain.

One feature to note is that here only the energies near the Fermi level are
shown. In our models, $A_{(E)}$ and $B_{(E)}$ also have extrema in other spectral
gaps, whenever the states are fairly localized.

\section{Discussion}

Here we offer some heuristic explanation of the last
two sections. In a-Si, there are two kinds of random potentials
leading to the localization of carriers:\ point defect (missing atoms or
impurities) and topological disorder. The former produces very localized
states which lie in the neighborhood of the Fermi level. These are bound
states around the point defects, and sit in the middle of gap. The
scattering suffered by the carriers is so strong that the localization
length of those states is smaller than the average bond length (for example,
the lone electron in dangling bond). \cite{ca} In contrast, topological disorder, the deviation of bond lengths and bond angles from their crystalline values, produces localized states primarily
lying in the tails of the valence band and the conduction band. The
localization length of those tail states may extend from one bond length to
a range including many atoms depending on the structure of the network.

Band tails exhibit localization because the density of states in their energy range is small, and there is less opportunity for delocalization via mixing, as discussed in the ``resonant cluster
proliferation" model. \cite{Resonance,Ludlam} In this picture, it is doubtful that a an initially localized energy state could persist in the midst of a continuum; it would inevitably mix with the reservoir of resonant states, if there was any overlap between the reservoir and localized state. Similarly, a well-localized state would survive if placed in a region of zero density of states. From this point of view, the band tails are simply between these limits.

According to the scattering theory of solids, \cite{ca,zim,wat} the
distortion of the atomic configuration from crystal mixes Bloch states with
different band indices and wave vectors. If these secondary scattering
components have large enough phase shifts relative to the primary Bloch wave
outside the distorted region, they will interfere destructively, because the
phase memory related to the primary Bloch wave has been lost. While inside the
distorted region, secondary scattering components are almost in-phase with
the primary Bloch wave (phase shift is less than $\pi$), and they will interfere
constructively inside this region. Thus if a distorted region can produce
scattered components with large momentum transfer, a localized state can be
formed by the interference of secondary components with the primary Bloch
wave.

With this mechanism, it is possible that several localized states have peaks
in the same distorted regions. This was observed in previous work \cite{Resonance} on M$_2$. We explored other models, M$_{4}$
and M$_{7}$, and found similar states with overlapping clusters in
space. On the other hand, one localized state may have several peaks in structurally distorted
regions which are spatially separated (Fig. \ref{F09}).

The observation that valence tail states are preferentially
localized on short bonds and conduction tail states localized on long
bonds (Fig. \ref{F10}) can be understood as follows. Short bond
length has two consequences: (1) it will increase the transition integral;
the energy of valence band states are lowered while the energy of conduction
states is lifted relative to a hypothetical reference crystal (diamond for a-Si);
(2)  Small bond lengths increase the charge at the center of the bonds
(states near valence edge). Thus if we make the significant assumption that the
valence and conduction tails should be associated with either short or long bonds, it
is clear that the electronic (band-energy) is optimized if the
valence tails are associated with short bonds, and the long bonds would ``do the least damage"
if put above the Fermi-level, or linked with the conduction tail. This completely neglects other
contributions to the total energy, though the observation of the effect in density functional
calculations suggests that the band energy {\it is} the key term.

\section{Conclusion}

We have discovered significant and reasonably model-independent features of
the network and electronic structure of a-Si. We believe that the results may prove relevant to
a number of other systems as well. \cite{jcp,swenson}

\section{Acknowledgements}

We thank the Army Research Office under MURI W91NF-06-2-0026, and the
National Science Foundation for support under grant No. DMR 0600073,
0605890. We acknowledge helpful conversations with T. Abtew and F. Inam.

\end{document}